\documentclass[3p,compress,numbers]{elsarticle}
\usepackage{epstopdf}
\usepackage{subfigure,graphicx,tabularx}
\usepackage{float}
\usepackage{amsmath, amsfonts, amssymb,mathrsfs}
\usepackage{amsthm}
\usepackage{epsf}
\usepackage{amsfonts}
\usepackage{stmaryrd}
\usepackage{amssymb}
\usepackage{leftidx}
\usepackage{color}
\usepackage{mathtools}
\usepackage{placeins}
\usepackage{booktabs}
\usepackage{enumitem}
\usepackage{caption}
\usepackage{multirow}
\usepackage{stackengine}
\usepackage[utf8]{inputenc}
\usepackage[english]{babel}
\usepackage{color}
\usepackage{pifont}
\usepackage{fontawesome}
\usepackage{hyperref}

\DeclareMathOperator{\sgn}{sgn}
\newcolumntype{?}{!{\vrule width 1pt}}

\theoremstyle{plain}
\newtheorem{theorem}{Theorem}[section]

\newtheorem{remark}[theorem]{Remark}

\biboptions{sort&compress}

\def\bfu{{\bf u}}

\def\bfw{{\bf w}}
\def\bfx{{\bf x}}

\def\bfE{{\bf E}}

\def\bfI{{\bf I}}

\def\bfN{{\bf N}}

\def\bfS{{\bf S}}

\def\bfX{{\bf X}}

\def\bfe{{\bf e}}

\def\bfw{\textbf{w}}

\def\l{\lambda}

\def\e0{\varepsilon_0}

\newcommand\sts{s_{\texttt{ts}}}
\newcommand\scs{s_{\texttt{cs}}}
\newcommand\shs{s_{\texttt{hs}}}
\newcommand\sbs{s_{\texttt{bs}}}
\newcommand\sss{s_{\texttt{ss}}}

\begin{document}

\begin{frontmatter}

\title{Abaqus implementation of the phase-field approach to fracture}

\author[Illinois]{Jiangwei Zhou}
\ead{jzhou106@illinois.edu}

\author[Abaqus]{Victor Lef\`evre}
\ead{Victor.LEFEVRE@3ds.com}

\author[Illinois]{Oscar Lopez-Pamies}
\ead{pamies@illinois.edu}

\address[Illinois]{Department of Civil and Environmental Engineering, University of Illinois, Urbana--Champaign, IL 61801, USA}

\address[Abaqus]{Dassault Syst\`emes, Providence, RI, USA}

\begin{abstract}

In this paper, we introduce an Abaqus user-subroutine implementation of the phase-field model of fracture initiated by Kumar, Francfort, and Lopez-Pamies (\emph{J. Mech. Phys. Solids} 112, 523--551, 2018). The implementation is presented for the basic case of isotropic linear elastic brittle materials with a Drucker-Prager strength surface. The material inputs are therefore the Young's modulus $E$, the Poisson's ratio $\nu$, the uniaxial tensile strength $\sts$, the uniaxial compressive strength $\scs$, and the fracture toughness $G_c$. The mathematical correspondence between the heat equation and the evolution equation for the phase field $v(\bfX,t)$ is exploited through a UMAT subroutine, wherein the driving force $c_{\texttt{e}}(\bfX,t)$ --- accounting for the material strength surface --- in the phase-field model plays the role of a heat source. To ensure that the equations for the displacement field $\bfu(\bfX,t)$ and the phase field $v(\bfX,t)$ are solved in a staggered manner --- a solution strategy that, as opposed to monolithic approaches, reliably yields the correct physical solution --- we make use of a combination of UEXTERNALDB, UAMP, MPC, and URDFIL subroutines. Within this approach, the MPC and URDFIL subroutines are additionally utilized to enforce the required bounding and irreversibility inequalities on $v(\bfX,t)$ via a primal-dual active set approach. The capabilities and robustness of the proposed implementation are demonstrated by simulating the ``Nine Circles of Elastic Brittle Fracture'' on a silicate glass --- a set of tests that span the entire range of well-settled experimental knowledge on fracture nucleation and propagation. These simulations also serve to illustrate the deployment of the code using several native Abaqus finite elements, including the simplicial elements CPS3, CAX3, and C3D4.

\vspace{0.2cm}

\endkeyword

\end{abstract}

\end{frontmatter}

\section{Introduction}

Over the past eight years, comparisons with a wide range of experimental observations \cite{KFLP18,KRLP18,KLP20,KBFLP20,KLP21,KRLP22,KLDLP24,KKLP24,KLP25,LPK25,KZDLP26,SRLP26,SMMRLP26,KBRLP26} on various nominally elastic-brittle materials --- both hard and soft --- have shown that the phase-field model initiated by Kumar, Francfort, and Lopez-Pamies \cite{KFLP18}, and developed further in \cite{FGLP19,KLP20,KBFLP20,KKLP24,LDLP24}, is able to describe and predict \emph{when} and \emph{where} cracks nucleate and propagate under monotonic, quasi-static, but otherwise arbitrary mechanical loads. Furthermore, numerical evidence indicates that this phase-field approach corresponds to a regularization of the sharp theory of fracture recently introduced in \cite{LPKFG2026}. This sharp theory describes the \emph{why} behind fracture nucleation and propagation in these materials: 
\smallskip

\noindent \emph{Cracks nucleate and propagate exclusively in regions where the strength surface of the material is exceeded, with their evolution dictated by minimizing the sum of the potential energy (the elastic energy minus the work done by externally applied forces) and surface energies.}

\smallskip

\noindent At present, the model is available in the open-source environments FEniCS\footnote{https://github.com/farhadkama/FEniCSx\_Kamarei\_Lopez-Pamies.} and RACCOON\footnote{https://github.com/hugary1995/raccoon.}  (within MOOSE). To make the model more widely accessible to the engineering community, this paper presents an Abaqus user-subroutine implementation. Attention is restricted to the basic case of isotropic linear elastic brittle materials with a Drucker-Prager strength surface.

As elaborated in Section \ref{Sec: Abaqus}, the implementation makes use of a novel combination of five main user subroutines. The first of these is a UMAT subroutine, wherein the mathematical correspondence between the heat equation native to Abaqus and the evolution equation for the phase field $v(\bfX,t)$ is exploited. This correspondence has already been utilized in implementations \cite{Azinpour2018,Paneda2021,Srivastava2026} of the initial phase-field models of fracture without strength \cite{Bourdin00}. Such early models lack a driving force $c_{\texttt{e}}(\bfX,t)$ in the evolution equation for $v(\bfX,t)$. It is this driving force that accounts for the strength surface of the material and, within the correspondence to the heat equation, plays the role of a heat source.

A second set of subroutines involving UEXTERNALDB, UAMP, MPC, and URDFIL is introduced to ensure that the equations for the displacement field $\bfu(\bfX,t)$ and the phase field $v(\bfX,t)$ are solved in a truly staggered, or alternating minimization, manner. This is crucial because such a solution strategy has been established to reliably yield the correct physical solution \cite{LDLP24,LPKFG2026}. We emphasize that the governing equations of phase-field fracture models can admit multiple solutions and that, as a result, the chosen numerical method directly impacts which solution is obtained; see, e.g., \cite{Knees2024}.

Finally, the MPC and URDFIL subroutines are further utilized to enforce the required inequalities on $v(\bfX,t)$ via a primal-dual active set approach. Namely, the phase field is constrained to satisfy the bounds $0 \leq v(\bfX,t) \leq 1$, alongside the irreversibility condition $v(\bfX,t_k) \leq v(\bfX,t_{k-1})$ whenever $v(\bfX,t_{k-1}) \leq v_\alpha$. Physically, this final constraint --- which dictates that the phase field cannot increase once it has decreased to a sufficiently small value $v_\alpha$ --- ensures that fracture is treated as an irreversible process.

Prior to presenting the Abaqus user-subroutine implementation, Sections \ref{Sec: The problem} and \ref{Sec: Discretizations} outline the governing equations for the phase-field model and their discretizations. In Section \ref{Sec: Nine Circles}, to demonstrate its capabilities and robustness, we confront the proposed implementation with the so-called ``Nine Circles of Elastic Brittle Fracture'' on a silicate glass \citep{KZDLP26}, a set of tests spanning the complete spectrum of well-settled experimental knowledge on fracture nucleation and propagation.

\section{The phase-field model}\label{Sec: The problem}

Consider a body that, initially, at time $t=0$, occupies the open bounded domain $\Omega$. We denote its boundary by $\partial\Omega$, with outward unit normal $\bfN$, and identify material points by their initial position vector $\bfX\in \Omega$.

The body is made of an isotropic linear elastic brittle material, characterized by an isotropic elastic energy density, Drucker-Prager strength surface, and scalar fracture toughness. Specifically, the elastic energy density is given by
\begin{equation}
W(\bfE)=\dfrac{E}{2(1+\nu)}\operatorname{tr}(\bfE^2)+\dfrac{E\nu}{2(1+\nu)(1-2\nu)}(\operatorname{tr}\bfE)^2,\label{Ingredient1 W}
\end{equation}
where $E$ and $\nu$ are the Young's modulus and Poisson's ratio, the Drucker-Prager strength surface reads
\begin{equation}
\mathcal{F}(\bfS)=\sqrt{J_2}+\dfrac{\scs-\sts}{\sqrt{3}\left(\scs+\sts\right)} I_1-\dfrac{2\scs \sts}{\sqrt{3}\left(\scs+\sts\right)}=0,\label{DP}
\end{equation}
where $\sts$ and $\scs$ are the uniaxial tensile and compressive strengths, respectively, and we write 
$$G_c$$
for the fracture toughness. In these expressions, $\bfE(\bfu)=\frac{1}{2}(\nabla\bfu+\nabla\bfu^T)$ denotes the symmetrized gradient of the displacement field $\bfu(\bfX,t)$, or infinitesimal strain tensor. Furthermore, $\bfS(\bfX,t)$ is the first Piola-Kirchhoff stress tensor, which in the present setting identifies with the Cauchy stress tensor, while $I_1={\rm tr}\,\bfS$ and $J_2=\frac{1}{2}\,{\rm tr}\,\bfS^2_{D}$ stand for the first and second principal invariants of $\bfS$ and $\bfS_D=\bfS-\frac{1}{3}({\rm tr}\,\bfS)\bfI$, respectively. In terms of the principal nominal stresses $s_1\geq s_2 \geq s_3$, these invariants read $I_1=s_1+s_2+s_3$ and $J_2=\frac{1}{3}(s^2_1+s^2_2+s^2_3-s_1 s_2-s_1 s_3-s_2 s_3)$.

The body is subjected to a time-dependent body force (per unit undeformed volume) $\mathbf{b}(\bfX,t)$ in $\Omega$, a time-dependent prescribed displacement $\overline{\bfu}(\bfX,t)$ on a part $\partial \Omega_{\mathcal{D}}$ of the boundary, and a time-dependent  surface traction (per unit undeformed area) $\overline{\mathbf{s}}(\bfX,t)$ on the complementary Neumann part $\partial\Omega_{\mathcal{N}}=\partial\Omega\setminus\partial \Omega_{\mathcal{D}}$.

In response to the applied body forces and boundary conditions --- all of which are assumed to be applied monotonically and quasi-statically in time --- the position vector $\bfX$ of a material point will move to a new position specified by $\bfx=\bfX+\bfu(\bfX,t)$. In addition to the deformation, the applied loading may result in the nucleation and subsequent propagation of cracks. We describe such cracks in a regularized fashion via the phase field $v=v(\bfX,t)$ taking values in the range $[0, 1]$. The value $v=1$ identifies the intact regions of the material and $v=0$ the regions that have been fractured. The transition from $v=1$ to $v=0$ is set to occur smoothly over regions of small thickness characterized by the regularization length  $\varepsilon>0$.

According to the phase-field fracture model put forth by Kumar, Francfort, and Lopez-Pamies \cite{KFLP18}, the displacement field $\bfu_k(\bfX)=\bfu(\bfX,t_k)$ and phase field $v_k(\bfX)=v(\bfX,t_k)$ at any material point $\bfX \in \overline{\Omega}=\Omega\cup\partial\Omega$ and at any given discrete time $t_k\in\{0=t_0,t_1,...,t_m,t_{m+1},...,t_M=T\}$ are determined by the system of coupled partial differential equations 
\begin{equation}
\left\{\begin{array}{ll}
{\rm Div}\left[v_{k}^2\dfrac{\partial W}{\partial \bfE}(\bfE(\bfu_{k}))\right]+v_{k}^2\textbf{b}(\bfX,t_k)=\textbf{0},& \,\bfX\in\Omega\vspace{0.2cm}\\
\bfu_k(\bfX)=\overline{\bfu}(\bfX,t_k), & \, \bfX\in\partial\Omega_{\mathcal{D}}\vspace{0.2cm}\\
\left[v_{k}^2\dfrac{\partial W}{\partial\bfE}(\bfE(\bfu_k))\right]\bfN=\overline{\textbf{s}}(\bfX,t_k),& \, \bfX\in\partial\Omega_{\mathcal{N}}
\end{array}\right. \label{BVP-u-theory}
\end{equation}
and
\begin{equation}
\left\{\begin{aligned}
& \text{Div} [\varepsilon \delta^\varepsilon G_c\nabla v_k] = \frac{8}{3}v_k W(\bfE(\bfu_k)) + \frac{4}{3}c_{\text{e}}(\bfX, t_k) - \frac{\delta^\varepsilon G_c}{2\varepsilon}, & & \text{if } \begin{cases} 0 < v_k(\bfX) < 1 & \text{for } v_{k-1}(\bfX) > v_\alpha \\ 0 < v_k(\bfX) < v_{k-1}(\bfX) & \text{for } v_{k-1}(\bfX) \le v_\alpha \end{cases}, & & \bfX \in \Omega \\[6pt]
& \text{Div} [\varepsilon \delta^\varepsilon G_c\nabla v_k] \ge \frac{8}{3}v_k W(\bfE(\bfu_k)) + \frac{4}{3}c_{\text{e}}(\bfX, t_k) - \frac{\delta^\varepsilon G_c}{2\varepsilon}, & & \text{if } \begin{cases} v_k(\bfX) = 1 \\ \text{or} \\ v_k(\bfX) = v_{k-1}(\bfX) \le v_\alpha \end{cases}, & & \bfX \in \Omega \\[6pt]
& \text{Div} [\varepsilon \delta^\varepsilon G_c\nabla v_k] \le \frac{8}{3}v_k W(\bfE(\bfu_k)) + \frac{4}{3}c_{\text{e}}(\bfX, t_k) - \frac{\delta^\varepsilon G_c}{2\varepsilon}, & & \text{if } v_k(\bfX) = 0 \text{ and } v_{k-1}(\bfX) > 0, & & \bfX \in \Omega \\[6pt]
& v_k(\bfX) = 0, & & \text{if } v_{k-1}(\bfX) = 0, & & \bfX \in \Omega \\[6pt]
& \nabla v_k \cdot \mathbf{N} = 0, & & & & \bfX \in \partial\Omega
\end{aligned}\right. \label{BVP-v-theory}
\end{equation}
subject to the initial conditions $\bfu(\bfX,0)\equiv \mathbf{0}$ and $v(\bfX,0)\equiv 1$, where $\nabla v_k=\nabla v(\bfX,t_k)$ and $v_\alpha$ is a sufficiently small threshold value below which the phase field is constrained to be non-increasing  ($v_k \leq v_{k-1}$) to enforce fracture irreversibility. In practice, it suffices to set $v_\alpha=0.05$. Furthermore, as elaborated in \cite{KBFLP20,KKLP24}, $c_\texttt{e}(\bfX,t)$ is a driving force whose specific constitutive prescription depends on the particular form of the strength surface $\mathcal{F}(\bfS)=0$, while $\delta^\varepsilon$ is a non-negative coefficient whose specific constitutive prescription depends in turn on the particular form of $c_\texttt{e}(\bfX,t)$; see also \cite{Chockalingam2025}. For the case of the Drucker-Prager strength surface (\ref{DP}) of interest here, the pertinent constitutive prescriptions read \citep{KKLP24}
\begin{align*}
c_{\texttt{e}}(\bfX,t)=\widehat{c}^\varepsilon_{\texttt{e}}(\bfE(\bfu),v)=v^2\beta^\varepsilon_2\sqrt{\mathcal{J}_2}+v^2\beta^\varepsilon_1 \mathcal{I}_1-v(1-\sgn(\mathcal{I}_1))W(\bfE(\bfu))
\end{align*}
and
\begin{align}
\delta^\varepsilon=\left(\dfrac{\sts+(1+2\sqrt{3})\,\shs}{(8+3\sqrt{3})\,\shs}\right)\dfrac{3 G_c}{16W_{\texttt{ts}}\varepsilon}+\dfrac{2}{5},\label{deltaeps}
\end{align}
where 
\begin{equation*}
\left\{\begin{array}{l}
\beta^\varepsilon_1=\dfrac{1}{\shs}\delta^\varepsilon\dfrac{G_c}{8\varepsilon}-\dfrac{2W_{\texttt{hs}}}{3\shs}\vspace{0.2cm}\\ \beta^\varepsilon_2=\dfrac{\sqrt{3}(3\shs-\sts)}{\shs\sts}\delta^\varepsilon\dfrac{G_c}{8\varepsilon}+
\dfrac{2W_{\texttt{hs}}}{\sqrt{3}\shs}-\dfrac{2\sqrt{3}W_{\texttt{ts}}}{\sts}
\end{array}\right. \quad {\rm and} \quad 
\left\{\begin{array}{l}
\mathcal{I}_1={\rm tr}\,\bfS=\dfrac{E}{1-2\nu}  {\rm tr}\,\bfE\vspace{0.2cm}\\
\mathcal{J}_2=\dfrac{1}{2}\,{\rm tr}\,\bfS^2_D=\dfrac{1}{2}\left(\dfrac{E}{1+\nu}\right)^2 {\rm tr}\,\bfE^2_D
\end{array}\right.
\end{equation*}
with $\bfE_D=\bfE-\frac{1}{3}({\rm tr}\,\bfE)\bfI$, $W_{\texttt{ts}}=\sts^2/(2E)$, $W_{\texttt{hs}}=3(1-2\nu)\shs^2/(2 E)$, and $\shs=2 \scs\sts/(3(\scs-\sts))$.

\begin{remark}
\emph{If the domain $\Omega$ occupied by the body contains a large pre-existing crack, the Neumann boundary condition (\ref{BVP-v-theory})$_5$ may be replaced with the Dirichlet boundary condition $v_k=0$ along the front and wake of the pre-existing crack.
}
\end{remark}

\begin{remark}\label{Remark: Staggered}
\emph{The governing equations (\ref{BVP-u-theory})-(\ref{BVP-v-theory}) are nonlinear and generally admit multiple solutions. Numerical evidence indicates that the correct physical solution --- namely, the solution of the sharp theory introduced in \cite{LPKFG2026} --- is reliably obtained using a staggered, or alternating minimization, scheme. In this approach, equation (\ref{BVP-u-theory}) is solved for the displacement field $\bfu_k$ at a fixed phase field $v_k$, and equation (\ref{BVP-v-theory}) is subsequently solved for $v_k$ with $\bfu_k$ fixed at its newly computed value. This process is iterated --- updating $\bfu_k$ and $v_k$ sequentially --- until both equations are satisfied within prescribed tolerances. Conversely, other approaches, such as monolithic schemes, may lead to non-physical solutions.
}
\end{remark}

\begin{remark}
\emph{The value of the regularization length $\varepsilon$ should be selected to be smaller than the smallest material length scale that arises from the dimensional interplay between elasticity (force/length$^2$), strength (force/length$^2$), and toughness (force/length) entering as material inputs in the governing equations (\ref{BVP-u-theory})-(\ref{BVP-v-theory}); see Remark 3.3 in \cite{LPKFG2026}. In practice, it typically suffices to set
\begin{equation}
\varepsilon< \dfrac{3 G_c}{16 W_{\texttt{ts}}}=\dfrac{3 E G_c}{8 \sts^2}.\label{eps}
\end{equation}
}
\end{remark}

\begin{remark}
\emph{In solving the governing equations (\ref{BVP-u-theory})-(\ref{BVP-v-theory}) via the finite element (FE) method, the mesh size $\texttt{h}$ must be sufficiently small to resolve the spatial variations of the phase field $v_k$ over the length scale $\varepsilon$. However, this still introduces a discretization error that scales with $\texttt{h}$. It is possible to include a correction in the formula (\ref{deltaeps}) for $\delta^\varepsilon$ so that the FE solutions of equations (\ref{BVP-u-theory})-(\ref{BVP-v-theory}) are consistent with the actual value $G_c$ of the fracture toughness of the material. For first-order simplicial FEs of size $\texttt{h}$, the formula for $\delta^\varepsilon$ with the correction reads \citep{KKLP24}
\begin{align}
\delta^\varepsilon=&\left(1+\dfrac{3}{8}\dfrac{\texttt{h}}{\varepsilon}\right)^{-2}\left(\dfrac{\sts+(1+2\sqrt{3})\,\shs}{(8+3\sqrt{3})\,\shs}\right)\dfrac{3 G_c}{16 W_{\texttt{ts}}\varepsilon}+\left(1+\dfrac{3}{8}\dfrac{\texttt{h}}{\varepsilon}\right)^{-1}\dfrac{2}{5}.
\label{delta-eps-final-h}
\end{align}
}
\end{remark}

\section{Weak form, FE discretization, and staggered solution scheme}\label{Sec: Discretizations}

\subsection{Weak form of the governing equations}\label{Sec: Weak}

By defining the admissible sets
\begin{equation*}
\mathcal{A}_{\bfu}=\left\{\bfu_k\in H^{1}(\Omega;\mathbb{R}^3): \bfu_k(\bfX)=\overline{\bfu}(\bfX,t_k),\;\bfX\in\partial\Omega_{\mathcal{D}}\right\}
\end{equation*}
and
\begin{equation*}
\mathcal{A}_v=\left\{v_k\in H^{1}(\Omega;\mathbb{R}): 0\leq v_k\leq 1 \text{ and } v_k\leq v_{k-1} \text{ for } v_{k-1}\leq v_\alpha \right\}
\end{equation*}
for the displacement field $\bfu_k$ and the phase field $v_k$, respectively, we can use standard calculations to reformulate the governing equations (\ref{BVP-u-theory})–(\ref{BVP-v-theory}) in weak form as the problem of finding the displacement field $\bfu_k\in \mathcal{A}_{\bfu}$ and phase field $v_k\in\mathcal{A}_v$ such that 
\begin{equation}\label{Weak-u-theory}
\displaystyle\int_{\Omega}\left(v_k^2\dfrac{\partial W}{\partial \bfE}(\bfE(\bfu_k))\right)\cdot\nabla\bfw\,{\rm d}\bfX-
\displaystyle\int_{\Omega}v_k^2\textbf{b}\cdot\bfw\,{\rm d}\bfX-
\displaystyle\int_{\partial\Omega_{\mathcal{N}}}\overline{\textbf{s}}\cdot\bfw\,{\rm d}\bfX=0\quad \forall \bfw\in \mathcal{A}_{\bfw},
\end{equation}
where $\mathcal{A}_{\bfw}=\{\bfw\in H^{1}(\Omega;\mathbb{R}^3): \bfw(\bfX)=\textbf{0},\;\bfX\in\partial\Omega_{\mathcal{D}}\}$, and
\begin{equation}\label{Weak-v-theory}
\displaystyle\int_{\Omega} \varepsilon \delta^\varepsilon G_c \nabla v_k \cdot \nabla (z - v_k) {\rm d}\bfX  + \displaystyle\int_{\Omega} \left( \frac{8}{3}v_k W(\bfE(\bfu_k)) + \frac{4}{3} c_{\texttt{e}}(\bfX, t_k) - \frac{\delta^\varepsilon G_c}{2\varepsilon} \right) (z - v_k)  {\rm d}\bfX \geq 0 \quad \forall z\in \mathcal{A}_z,
\end{equation}
where $\mathcal{A}_{z} = \{ z \in H^{1}(\Omega;\mathbb{R}) : 0 \le z(\bfX) \le v_{k-1}(\bfX) \text{ if } v_{k-1}(\bfX) \le v_{\alpha}\text{, else } 0 \le z(\bfX) \le 1 \}$.

\subsection{FE discretization}\label{Sec: Space disc}

With the weak forms (\ref{Weak-u-theory}) and (\ref{Weak-v-theory}) established, we proceed to discretize them via a standard Galerkin FE method. Let the domain $\Omega$ be partitioned into a conforming discretization of finite elements. The displacement trial field $^{\texttt{h}}\bfu_k(\bfX)$ and the phase trial field $^{\texttt{h}}v_k(\bfX)$ in (\ref{Weak-u-theory}) and (\ref{Weak-v-theory}) can then be written in the global form
\begin{equation}\label{FE-approximation}
^{\texttt{h}}u_{k,i}(\bfX) = \sum_{n=1}^{\texttt{N}_n} N^{(n)}(\bfX) u_{k,i}^{(n)} \quad \text{and} \quad {}^\texttt{h} v_k(\bfX) = \sum_{n=1}^{\texttt{N}_n} N^{(n)}(\bfX) v_{k}^{(n)},
\end{equation}
where $N^{(n)}(\bfX)$ stands for the global shape function associated with node $n$, $\texttt{N}_n$ is the total number of nodes, while $u_{k,i}^{(n)}$ and $v_{k}^{(n)}$ correspond, respectively, to the component $i$ of the displacement field $^{\texttt{h}}\bfu_k(\bfX)$ and the phase field $^{\texttt{h}}v_k(\bfX)$ at node $n$ and time $t_k$.

Upon substitution of (\ref{FE-approximation}) in (\ref{Weak-u-theory}) and (\ref{Weak-v-theory}), the resulting residual for the displacement component $i$ at node $n$ can be written as
\begin{equation}\label{Ru}
\mathcal{R}_{u,i}^{(n)} = \displaystyle\int_\Omega \left( {}^{\texttt{h}}v_k \right)^2 \frac{\partial W}{\partial E_{ij}}\left( \mathbf{E}({}^{\texttt{h}}\mathbf{u}_k) \right) \frac{\partial N^{(n)}}{\partial X_j} \mathrm{d}\bfX- \int_\Omega \left( {}^{\texttt{h}}v_k \right)^2 b_i N^{(n)} \mathrm{d}\bfX - \int_{\partial\Omega_{\mathcal{N}}} \overline{s}_i N^{(n)} \mathrm{d}\bfX = 0,
\end{equation}
while the residual for the phase field at node $n$ reads
\begin{equation}\label{Rv-1}
\sum_{n=1}^{\texttt{N}_n} \mathcal{R}_v^{(n)} \left( z^{(n)} - v_k^{(n)} \right) \geq 0 \quad \forall z^{(n)} \in \mathcal{A}_z^{\texttt{h}},
\end{equation}
where $\mathcal{A}_z^{\texttt{h}}=\{ z^{(n)} : 0 \le z^{(n)} \le v_{k-1}^{(n)} \text{ if } v_{k-1}^{(n)} \le v_{\alpha}\text{, else } 0 \le z^{(n)} \le 1 \text{ for } n=1,\dots,\texttt{N}_{n} \}$ and
\begin{equation}\label{Rv-2}
\mathcal{R}_v^{(n)} = \int_\Omega \varepsilon \delta^\varepsilon G_c \nabla {}^{\texttt{h}}v_k \cdot \nabla N^{(n)} \mathrm{d}\bfX + \int_\Omega \left( \frac{8}{3} {}^{\texttt{h}}v_k W \big( \mathbf{E}({}^{\texttt{h}}\mathbf{u}_k) \big) + \frac{4}{3} c_{\mathrm{e}}(\bfX, t_k) - \frac{\delta^\varepsilon G_c}{2\varepsilon} \right) N^{(n)} \mathrm{d}\bfX.
\end{equation}

With the help of the definition
\[
U_n = 
\begin{cases} 
v_{k-1}^{(n)} & \text{if } v_{k-1}^{(n)} \leq v_\alpha \\ 
1 & \text{otherwise} 
\end{cases}
\]
for the upper bound for the admissible phase field at node $n$, note that the variational inequality (\ref{Rv-1}) with (\ref{Rv-2}) dictates three exhaustive possible states for the phase field $v_k^{(n)}$ at any given node $n$:
\begin{itemize}
    \item \emph{Case 1: Stationary state (upper bound).} When the phase field remains at its maximum admissible value, $v_k^{(n)} = U_n$. In this state, any admissible test value $z^{(n)}$ satisfies $\left( z^{(n)} - v_k^{(n)} \right) \leq 0$. To satisfy the variational inequality, the nodal residual must be non-positive:
    \[
    \mathcal{R}_v^{(n)} \leq 0.
    \]

    \item \emph{Case 2: Active fracture evolution (interior).} When the material is actively fracturing but not yet fully broken, the phase field is strictly within the interior bounds: $0 < v_k^{(n)} < U_n$. Because $z^{(n)}$ can be chosen to be larger or smaller than $v_k^{(n)}$, the perturbation $\left( z^{(n)} - v_k^{(n)} \right)$ can take both positive and negative values. Consequently, the variational inequality requires the residual to be exactly zero:
    \[
    \mathcal{R}_v^{(n)} = 0.
    \]

    \item \emph{Case 3: Fully fractured state (lower bound).} When the material is fully fractured, $v_k^{(n)} = 0$. In this state, any admissible test value $z^{(n)}$ is strictly non-negative, meaning $\left( z^{(n)} - v_k^{(n)} \right) \geq 0$. To satisfy the variational inequality, the nodal residual must be non-negative:
    \[
    \mathcal{R}_v^{(n)} \geq 0.
    \]
\end{itemize}

\subsection{Staggered solution scheme}

In view of Remark \ref{Remark: Staggered}, the discretized coupled system of equations (\ref{Ru}) and (\ref{Rv-1})-(\ref{Rv-2}) for the degrees of freedom of the displacement ($u_{k,i}^{(n)}$) and phase field ($v_k^{(n)}$) is to be solved in a staggered manner at each time step $t_k$. Specifically, the system is to be solved iteratively by holding the phase field $^{\texttt{h}}v_k$ constant while solving equation (\ref{Ru}) for the updated displacement field $^{\texttt{h}}\bfu_k$, followed by freezing the displacement field to solve the constrained variational inequality (\ref{Rv-1}), with (\ref{Rv-2}), for the phase field. This alternating update loop is repeated until convergence is achieved at a constant load level.

\section{The Abaqus user-subroutine implementation}\label{Sec: Abaqus}

To solve in a staggered manner in Abaqus the coupled system of discretized governing equations (\ref{Ru}) and (\ref{Rv-1})-(\ref{Rv-2}) for the degrees of freedom that define the FE approximations (\ref{FE-approximation}) for the displacement and phase fields, we use an implementation involving five main user subroutines: UMAT, UEXTERNALDB, UAMP, MPC, and URDFIL. The code for this implementation has been made available on GitHub,\footnote{Links to the GitHub repositories are available at \url{https://pamies.cee.illinois.edu/repositories/}.} alongside the input files of the examples presented in Section \ref{Sec: Nine Circles}. In the sequel, we outline the specific role and integration of each subroutine within this staggered implementation. In particular, Section \ref{Sec: UMAT} outlines the evaluation of the constitutive response, Section \ref{Sec: Staggered} details the staggered solution scheme, Section \ref{Sec: Inequalities}, the enforcement of the phase-field inequalities, and Section \ref{Sec: Acceptance} presents the load-level acceptance criteria. The setup and output of a simulation are described in Sections \ref{Sec: User} and \ref{Sec: Output}.

\subsection{The \emph{UMAT} subroutine}\label{Sec: UMAT}

The constitutive response and the source terms in the discretized displacement and phase-field governing equations (\ref{Ru}) and (\ref{Rv-1})-(\ref{Rv-2}) are evaluated at every integration point via a UMAT subroutine. Within this subroutine, the phase-field variable $v_k$ is mapped to the temperature degree of freedom native to Abaqus. The intact elastic stiffness matrix is multiplied by the degradation function $v_k^2$, yielding the degraded tangent stiffness (\texttt{DDSDDE}) and stress (\texttt{STRESS}) tensors required to solve for the displacement degrees of freedom. The UMAT also computes the phase-field source term and its tangent derivative. In the native Abaqus coupled thermomechanical framework, these phase-field terms correspond to the volumetric heat source (\texttt{RPL}) and its thermal tangent (\texttt{DRPLDT}), which are required to numerically solve for the phase-field degrees of freedom. Crucially, the UMAT evaluates these terms without incorporating penalty parameters. Instead, the exact enforcement of the phase-field inequality constraints (\ref{Rv-1}) is handled globally via multi-point constraints. Furthermore, cross-terms in the stiffness matrix are not required thanks to the staggered nature of the solution scheme.

\subsection{The staggered solution scheme via \emph{UEXTERNALDB}, \emph{UAMP}, \emph{MPC}, and \emph{URDFIL} subroutines}\label{Sec: Staggered}

The staggered solver is implemented using a combination of four different Abaqus user subroutines. In a nutshell, multi-point constraints are used to selectively freeze either the displacement field or the phase field, while the other field is solved for during an Abaqus increment. Unlike standard Abaqus simulations where the load level updates after each successful increment, the load level here (\texttt{lambda}) is held fixed until the staggered acceptance criteria are satisfied. This yields a sequence of increments at a constant load level that alternately solves for one field while the other remains fixed. Note that because Abaqus advances its internal time with every increment while \texttt{lambda} is held constant, the native Abaqus time loses its physical meaning. Instead, the load parameter \texttt{lambda} serves as the physical time.

Specifically, at the beginning of the analysis, the UEXTERNALDB subroutine initializes the module state and mutexes, enabling the subroutines to access the global state variables controlling the staggered scheme. During each increment, the load level is enforced via the UAMP subroutine, which can then be used by other components --- such as the DISP subroutine --- to scale and impose boundary conditions. The MPC subroutine freezes the inactive degrees of freedom for the current solve increment by selectively activating the \texttt{lmpc} flag, ensuring the active field is solved while the other remains frozen.

Finally, the URDFIL subroutine functions as the master scheduler. It runs after each converged increment and dictates how the simulation proceeds:
\begin{itemize}
\item a displacement solve increment is always followed by a phase-field solve increment at the same load level;
\item a phase-field solve increment is followed by another displacement solve increment at the same load level, unless the load level acceptance criteria are met;
\item if the acceptance criteria are met, the converged fields are accepted, the load level is updated, and a new staggered pass --- a displacement solve increment followed by a phase-field solve increment --- begins at the new load level.
\end{itemize}
If the system diverges, URDFIL reduces the load-level increment and reattempts the staggered solution at a smaller load step.

Furthermore, URDFIL ensures that only accepted fields are written to the result databases for a given load level. Writing intermediate results at the end of every increment, as would natively occur in Abaqus, lacks physical meaning in this context and would needlessly inflate the size of the result databases.

At the end of the analysis, UEXTERNALDB performs a final audit to confirm the target load level was successfully reached and the corresponding results were written to the database.

\subsection{Enforcement of phase-field inequalities via \emph{MPC} and \emph{URDFIL} subroutines: A primal-dual active set approach}\label{Sec: Inequalities}

The MPC and URDFIL subroutines also enforce the phase-field bounds $0 \le v_k \le 1$ and $0 \le v_k \le v_{k-1}$ if $v_{k-1}\le v_\alpha$ through a primal-dual active set approach. In this approach, the active set corresponds to all the nodes whose phase-field degree of freedom satisfies 

\begin{equation}\label{v_equalities}
\left\{\begin{array}{ll}
v_k =0\quad \text{or}\quad v_k =1 & \text{if } v_{k-1}>v_\alpha \\
v_k =0\quad \text{or}\quad v_k =v_{k-1} & \text{if } v_{k-1}\le v_\alpha
\end{array}
\right..
\end{equation}
Every phase-field degree of freedom requiring a solution is linked to a reference node via a constraint equation. This formulation ensures the phase field remains fixed when frozen, exploiting the fact that the degree of freedom of the reference node is permanently set to zero.

During a phase-field increment, the MPC subroutine pins the phase-field degrees of freedom for nodes within the active set exactly at their upper irreversibility bound or lower zero bound (\ref{v_equalities}). Once the unconstrained phase-field degrees of freedom are solved, URDFIL updates the active set by performing the following primal and dual checks:
\begin{itemize}
\item \emph{Primal check (pinning).} The newly computed phase-field values at free nodes are checked against their admissible bounds. If a value exceeds its upper bound ($v_k > 1$ or $v_k > v_{k-1}$) or falls below zero ($v_k < 0$), both within a tolerance \texttt{pin\_tol}, its node is added to the active set and pinned to that respective bound for the next staggered pass.

\item \emph{Dual check (release).} For each pinned node, URDFIL reads the reaction force at its associated reference node, which corresponds to the discrete Lagrange multiplier $\mathcal{R}_v^{(n)}$ defined by (\ref{Rv-2}). If $\mathcal{R}_v^{(n)} \ge 0$ at an upper-pinned node ($v_k = 1$ when $v_{k-1} > v_\alpha$, or $v_k = v_{k-1}$ when $v_{k-1} \le v_\alpha$), the node is removed from the active set and released. Similarly, if $\mathcal{R}_v^{(n)} \le 0$ at a lower-pinned node ($v_k = 0$), it is also removed from the active set and released.
    
    The dual checks on the sign of $\mathcal{R}_v^{(n)}$ are evaluated against the dual tolerance $\texttt{dual\_tol}=\max\{$ $\texttt{dual\_abs\_tol},$ $ \texttt{dual\_rel\_tol} * \texttt{rf\_scale0}\}$, where \texttt{rf\_scale0} is the model's characteristic dual magnitude --- the largest reaction-force magnitude over the phase-field reference nodes --- determined once based on the first accepted displacement field and held for the rest of the analysis.
         
\end{itemize}

Once the acceptance criteria are met, the irreversibility upper bound is updated to match the newly accepted phase-field values exclusively for the nodes that have reached or fallen below the threshold $v_\alpha$. For all other nodes, the upper bound resets to 1, leaving their degrees of freedom reversible. 

\subsection{The acceptance criteria}\label{Sec: Acceptance}

A load level is accepted only when two conditions are fulfilled simultaneously: ($i$) the displacement and phase fields satisfy the staggered convergence criteria, and ($ii$) the active set achieves a fixed point.

Rather than evaluating standard solver residuals, the staggered solver monitors the incremental change of the fields across staggered solve passes. Four convergence criteria are available: \texttt{V\_NORM} (relative $L^2$ norm of the phase-field increment, $\parallel\Delta v\parallel_2/\parallel v\parallel_{2}$), \texttt{E\_NORM} (relative $L^2$ norm of the strain increment, $\parallel\Delta \bfE\parallel_2/\parallel\bfE\parallel_2$), \texttt{U\_NORM} (relative $L^2$ norm of the nodal displacement increment, $\parallel\Delta \bfu\parallel_2/\parallel \bfu\parallel_{2}$), and \texttt{DV\_INF\_NORM} (absolute $L^\infty$ norm of the phase-field increment, $\parallel\Delta v\parallel_\infty=\max_i|\Delta v_i|$). Each criterion can be defined with its own tolerance, in which case, staggered convergence is achieved when every convergence criterion is satisfied simultaneously. Staggered convergence is also considered to be achieved when the number of staggered passes reaches the user-defined maximum number of staggered passes \texttt{nmax\_passes}; typically, it suffices to set \texttt{nmax\_passes}$ = 100$.

The active set is considered to have reached a fixed point when no nodes were added to or released from the active set during the last staggered pass. This condition can be relaxed to allow at most \texttt{nmax\_vflips} changes during the last staggered pass.

\subsection{The setup of a simulation}\label{Sec: User}

To set up a simulation, the user must prescribe a set of material constants and computational parameters, provide model and mesh information, as well as specify key numerical controls using Abaqus parameter tables. A preprocessing Python script is available to incorporate all of this information into existing Abaqus input files.

\subsubsection{Material constants and computational parameters} 

As noted in Section \ref{Sec: The problem}, the material constants governing the elastic deformation and fracture of an isotropic linear elastic material with a Drucker-Prager strength surface are Young's modulus $E$, Poisson's ratio $\nu$, the uniaxial tensile strength $\sts$, the uniaxial compressive strength $\scs$, and the fracture toughness $G_c$. These five constants, together with the regularization length $\varepsilon$ --- which must satisfy inequality (\ref{eps}) --- are prescribed for each material definition via the \texttt{KFLP\_MATERIAL} parameter table. Consequently, this table requires exactly six floating-point numbers and one integer. The integer input, \texttt{mesh\_corr}, allows the user to select formula (\ref{delta-eps-final-h}) --- instead of (\ref{deltaeps}) --- for the coefficient $\delta^\varepsilon$ to account for the FE discretization.

\subsubsection{Model and mesh information} 

The code requires a \texttt{STAG\_SIZES} table, which defines the size of the model. This table must contain a single row with four or five integers: the total number of nodes in the model (\texttt{nall}), the number of phase-field constraints (\texttt{nv}), and the number of constraints for each of the displacement components (\texttt{nu1}, \texttt{nu2}, and optionally \texttt{nu3}). For models defined in terms of part instances, a \texttt{STAG\_INSTANCES} table must list the instance names, one per row. 

\subsubsection{Algorithmic choices} 

Global algorithmic controls can be defined via the \texttt{STAG\_ALG\_CONTROLS} table. This table allows the user to prescribe up to six operational controls. These include the maximum number of staggered passes per load level (\texttt{nmax\_passes}), a default staggered convergence tolerance (\texttt{stag\_tol}), and the irreversibility threshold (\texttt{v\_alpha}). Furthermore, this table incorporates several advanced controls:

\begin{itemize}
    \item\texttt{unconverged\_policy}. Determines whether the analysis terminates with an error or commits the unconverged state when the acceptance conditions remain unsatisfied at the minimum load increment.
    \item \texttt{report\_mode}. Determines whether an ODB frame is published per accepted load level or per accepted increment.
    \item\texttt{verbose}. Determines whether informational messages, including the echo of the prescribed settings and material constants, are written to the message (\texttt{.msg}) file in addition to errors and warnings.
\end{itemize}

Convergence of the staggered scheme can be controlled by the user-defined \texttt{STAG\_CONV\_CRITERIA} table. Each line of the table must contain the name of the criterion (one of \texttt{V\_NORM}, \texttt{E\_NORM}, \texttt{U\_NORM}, \texttt{DV\_INF\_NORM} as described in Section \ref{Sec: Acceptance}) and the corresponding tolerance.

The controls for the primal-dual active set approach to enforce the phase-field inequalities can be adjusted via the user-defined \texttt{STAG\_ACTIVE\_SET} table. This table allows the user to prescribe up to six operational controls. They include the number of active set changes allowed in the last staggered solve pass \texttt{nmax\_vflips}, the primal pin tolerance \texttt{pin\_tol} that dictates how tightly the multi-point constraints hold the phase field at its bounds, the absolute floor of the dual tolerance \texttt{dual\_abs\_tol}, and the relative factor of the dual tolerance \texttt{dual\_rel\_tol}. This table also includes the advanced controls:
\begin{itemize}
    \item\texttt{release\_mode}. Controls whether the dual check may release a pinned node after the displacement
solve, or exclusively after the phase-field solve.
    \item \texttt{vproj\_mode}. Selects when the phase field is projected onto its bounding box: during every Newton
iteration, only during increments re-attempted by Abaqus following non-convergence, or never.
\end{itemize}

\subsubsection{Step and amplitude definitions}

As mentioned in Section \ref{Sec: Staggered}, the staggered solution scheme decouples the physical evolution of the solution in time from Abaqus' internal time. As a result, standard time-dependent features --- such as step-time-based load ramping or conventional tabular amplitudes --- cannot be used directly.\footnote{Because \texttt{lambda} acts as the effective physical time, any future extension of the \texttt{UMAT} subroutine to rate- or time-dependent constitutive models must integrate rate equations using physical time increments derived from $\texttt{dlambda}$ rather than Abaqus' native increment \texttt{DTIME}.} Instead, steps and amplitudes are parameterized directly in terms of the staggered load level $\texttt{lambda}$ using parameter tables.

The mandatory \texttt{STAG\_STEPS} table configures the load increments for each \texttt{*Step} definition, mirroring the \texttt{*Static} procedure definition line. Each row defines four parameters: the initial load-level increment (\texttt{dlam\_init}), the target load-level up to which the step advances from zero (\texttt{lam\_target}), the minimum allowable load-level increment during automatic cutbacks (\texttt{dlam\_min}), and the maximum allowable load-level increment (\texttt{dlam\_max}).

\subsection{Output}\label{Sec: Output}

The user can request to output a strength-surface violation indicator, \texttt{SDV\_SSV\_IND}, which is stored in the state variable \texttt{SDV1}. Rather than simply acting as a binary flag for whether the stress $\bfS = \partial W(\bfE(\bfu_k))/\partial\bfE$ exceeds the strength surface --- which, again, indicates regions where cracks are allowed to nucleate and propagate --- \texttt{SDV\_SSV\_IND} also specifies the number of strictly positive principal stresses during a violation. This provides insight into how the strength surface is violated in stress space. Specifically, \texttt{SDV\_SSV\_IND} is defined as
\begin{equation*}
\texttt{SDV\_SSV\_IND} = \begin{cases}
-1 & \text{if } \mathcal{F}(\bfS) < 0 \text{ (strength surface not violated)} \\[4pt]
0  & \text{if } \mathcal{F}(\bfS) \ge 0 \text{ with } s_1 \le 0,\, s_2 \le 0,\, s_3 \le 0 \\[4pt]
1  & \text{if } \mathcal{F}(\bfS) \ge 0 \text{ with } s_1 > 0,\, s_2 \le 0,\, s_3 \le 0 \\[4pt]
2  & \text{if } \mathcal{F}(\bfS) \ge 0 \text{ with } s_1 > 0,\, s_2 > 0,\, s_3 \le 0 \\[4pt]
3  & \text{if } \mathcal{F}(\bfS) \ge 0 \text{ with } s_1 > 0,\, s_2 > 0,\, s_3 > 0
\end{cases}.
\end{equation*}
The Appendix illustrates the \texttt{SDV\_SSV\_IND} evolution for the indentation problem presented in Section~\ref{Sec: Indentation}.

Finally, a Python script is available to postprocess the raw Abaqus output database, which replaces the native Abaqus time by the load parameter and physical time \texttt{lambda}.  

\section{Compliance with ``The Nine Circles of Elastic Brittle Fracture''}\label{Sec: Nine Circles}

In a recent contribution, Kamarei et al. \cite{KZDLP26} introduced ``The Nine Circles of Elastic Brittle Fracture,'' a suite of nine benchmark problems designed to critically evaluate the validity of prospective fracture theories. As defining features, these benchmarks can be carried out experimentally with standard testing equipment and, more critically, span the full spectrum of well-established experimental knowledge for both fracture nucleation and propagation in both hard and soft elastic brittle materials. Specifically, they probe fracture nucleation dominated by material strength under uniform stresses (circles 1 through 3), by the Griffith energy competition from large pre-existing cracks (circle 4), and by the more complex interplay between strength and Griffith energy competition under arbitrary non-uniform stress states (circles 5 through 7). For propagation, the suite examines crack growth under opening (circle 8) and tearing (circle 9) conditions. Failure to accurately resolve even a single ``circle'' disqualifies a theory as a viable framework for describing, and hence predicting, fracture in general.

\begin{table}[H]\centering\small
\caption{Material constants for silicate glass.}
\begin{tabular}{l?cc}
\toprule
Elasticity constants &  $E$ (GPa) &  $\nu$  \\
\midrule
                     &  $70$      & 0.22     \\
\midrule
\midrule
Strength constants   & $s_{\texttt{ts}}$ (MPa) & $s_{\texttt{cs}}$ (MPa) \\
\midrule
                     & $40$                  & $1000$                 \\
\midrule
\midrule
Fracture toughness   & $G_c$ (N/m) & \\
\midrule
                               & $10$ & \\
\bottomrule
\end{tabular} \label{Table1}
\end{table}
In their work, Kamarei et al.~\cite{KZDLP26} showed that the available FEniCS and RACCOON implementations of the phase-field model (\ref{BVP-u-theory})-(\ref{BVP-v-theory}) successfully pass all nine circles. In this section, we deploy the proposed Abaqus user-subroutine implementation to simulate these benchmarks. Because this implementation is specific to linear elastic brittle materials, we focus exclusively on the seven circles (circles 1 through 6 and 8) pertaining to hard materials. As in \cite{KZDLP26}, the simulations are carried out for a silicate glass with elastic energy density (\ref{Ingredient1 W}), Drucker-Prager strength surface (\ref{DP}), scalar fracture toughness $G_c$, and the material constants listed in Table~\ref{Table1}. 

According to the values in Table~\ref{Table1}, it follows from (\ref{eps}) that the regularization length must satisfy $\varepsilon<0.164$~mm. The simulations presented in this section use $\varepsilon=0.02,0.075, 0.10$, and 0.15~mm and are generated with unstructured meshes of maximum element size $\texttt{h}\leq\varepsilon/3$ in the regions where fracture nucleation and propagation take place.

For circles 1 through 3, the stress fields are essentially uniform prior to fracture nucleation. Following previous works \cite{KFLP18,KBLP25}, we introduce a random variation of up to $\pm 10\%$ into the uniaxial tensile strength $\sts$ across the specimens in those problems. This small perturbation is applied over subregions of size 0.40~mm, which is sufficient to break symmetry and trigger fracture nucleation at single sites; see also \cite{Guilleminot2024}. Realizations of such a stochastic distribution are shown in Figs.~\ref{Fig1}(b), \ref{Fig3}(b), and~\ref{Fig5}(b) for circles 1, 2, and 3, respectively. 

Finally, we note that circles 1, 2, 4, 5, and 8 are simulated under plane-stress conditions using the 3-node linear triangular CPS3 elements in Abaqus. On the other hand, the results for circles 3 and 6 are generated using the 3D and axisymmetric elements C3D4 and CAX3, respectively.

\subsection{Fracture nucleation dominated by strength}

\subsubsection{Uniaxial tension test}

%
\begin{figure}[b!]
\centering\includegraphics[width=0.9\linewidth]{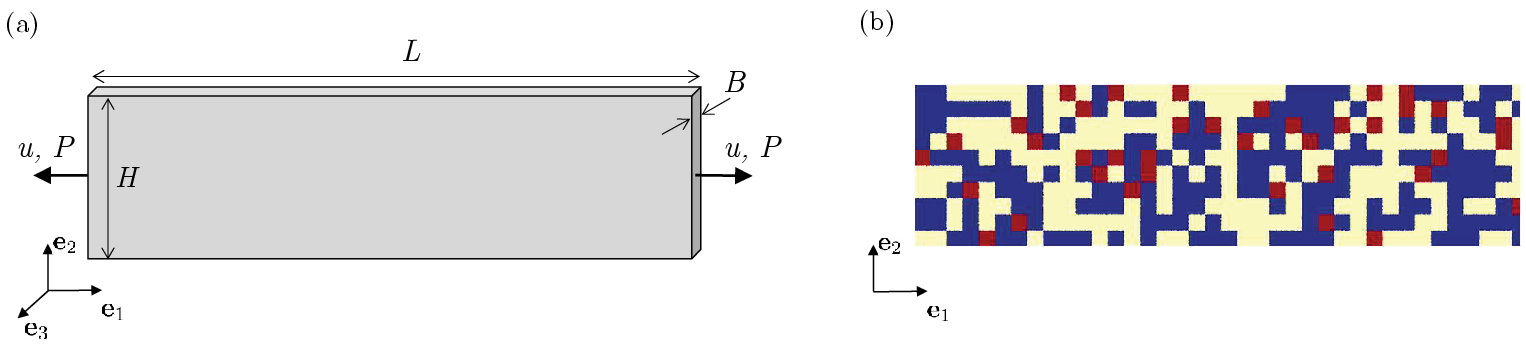}
\caption{\small (a) Schematic of the uniaxial tension test. The specimen dimensions are $L=15$ mm, $H=4$ mm, and $B=0.5$ mm. (b) A realization of the subregions of size $0.40$ mm over which the uniaxial tensile strength $\sts$ of the material is randomly perturbed by $\pm 2\%$ (red $=0.98\sts$, yellow $=\sts$, blue $=1.02\sts$).}\label{Fig1}
\end{figure}
%
%
\begin{figure}[b!]
\centering\includegraphics[width=0.9\linewidth]{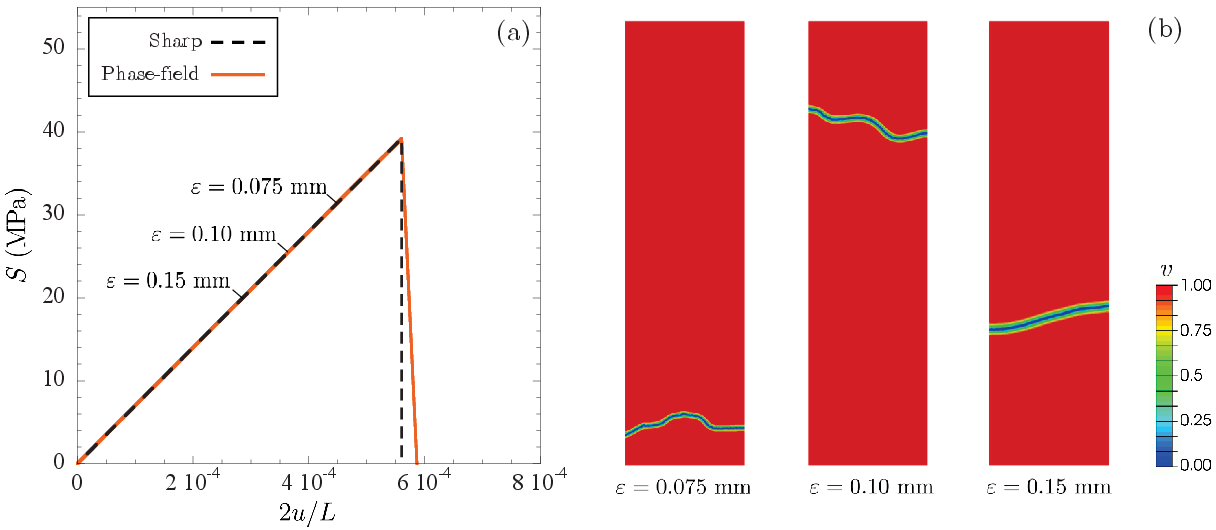}
\caption{\small (a) Comparisons between the predictions by the phase-field model for the stress-strain response of the specimen and the reference result (\ref{Response-Uni}) for three different values of the regularization length $\varepsilon$. (b) Contour plots of the phase field $v$ over the undeformed configuration of the specimen right after fracture nucleation.}\label{Fig2}
\end{figure}
%

We begin by considering the $1^{\text{st}}$ circle, a prismatic specimen of length $L=15$ mm, width $H=4$ mm, and thickness $B=0.5$ mm that is subjected to uniaxial tension at its ends by the application of a displacement $u$. The specimen undergoes a uniform axial strain of $2u/L$ and a corresponding uniform uniaxial stress $S=P/(H B)$, where $P$ is the resultant force at the ends of the specimen. In the Cartesian laboratory frame of reference indicated in Fig.~\ref{Fig1}(a),
\begin{equation*}
\bfE = \dfrac{2u}{L}\bfe_1\otimes\bfe_1 + e_{t}(\bfe_2\otimes\bfe_2 + \bfe_3\otimes\bfe_3) \quad \text{and} \quad \bfS = S\bfe_1\otimes\bfe_1,
\end{equation*}
where $e_t$ is the transverse strain due to the Poisson effect. This uniform state of stress and strain continues until the stress $S$ reaches the uniaxial tensile strength $\sts$ of the material. At this threshold, the abrupt nucleation/propagation of a crack perpendicular to the loading axis severs the specimen in two. Consequently, the measures of stress $S$ and strain $2u/L$ are related according to
%
\begin{equation}
S = \begin{cases} 
E\dfrac{2u}{L} & \text{if } 0 \leq u < u_{\texttt{ts}} \\[8pt]
0 & \text{if } u_{\texttt{ts}} \leq u 
\end{cases}, \label{Response-Uni}
\end{equation}
%
%
where $u_{\texttt{ts}}$ denotes the value of the displacement at which $S=\sts$. Note that the exact location of the severing crack is arbitrary, reflecting the inherently stochastic nature of macroscopic material strength.

Figure~\ref{Fig2} compares the predictions generated by the phase-field model with the reference result --- labeled \emph{sharp} in the plot due to its sharp description of fracture. Specifically, part~(a) compares the stress-strain response ($S$ vs.~$2u/L$) for three values of the regularization length $\varepsilon$, while part~(b) presents contour plots of the phase field $v$ over the undeformed configuration of the specimen right after fracture nucleation. The main observation from these results is that, irrespective of the value of $\varepsilon$, the phase-field model accurately predicts the reference result (\ref{Response-Uni}), as well as the associated abrupt nucleation/propagation of a severing crack with the correct orientation.

\subsubsection{Biaxial tension test}

%
\begin{figure}[b!]
\centering\includegraphics[width=0.8\linewidth]{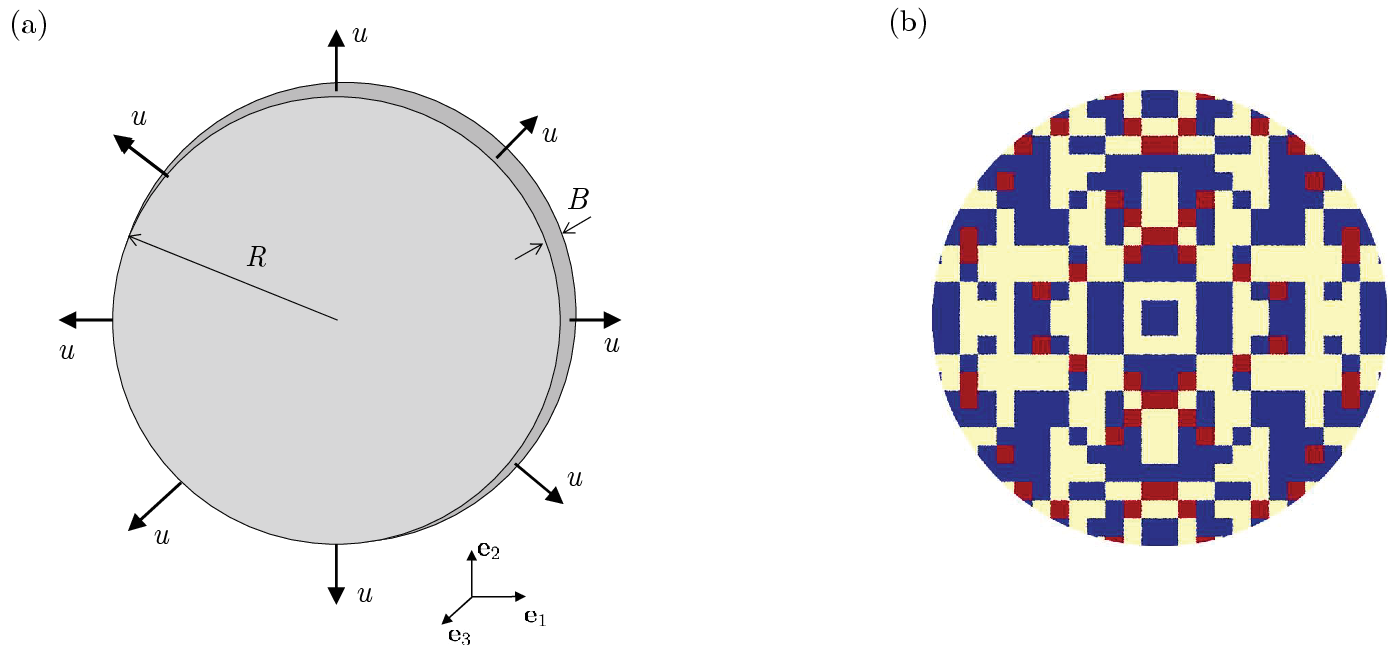}
\caption{\small (a) Schematic of the biaxial tension test. The specimen dimensions are $R=5$ mm and $B=0.25$ mm. (b) A realization of the subregions of size $0.40$ mm over which the uniaxial tensile strength $\sts$ of the material is randomly perturbed.}\label{Fig3}
\end{figure}
%
%
\begin{figure}[t!]
\centering\includegraphics[width=0.9\linewidth]{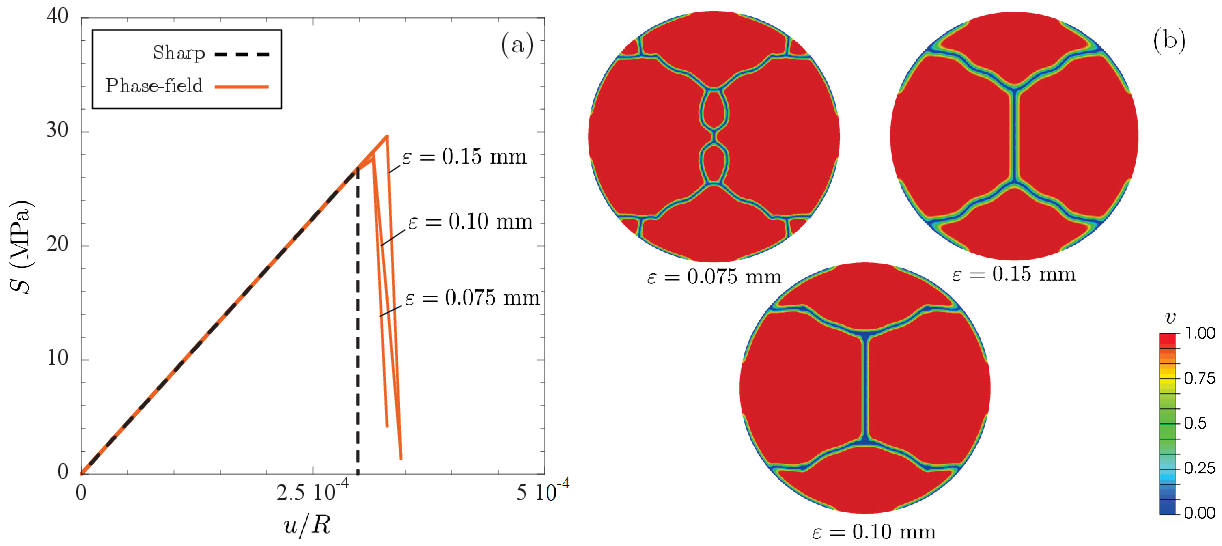}
\caption{\small (a) Comparisons between the predictions by the phase-field model for the stress-strain response of the plate and the reference result (\ref{Response-Bi-Lin}) for three different values of the regularization length $\varepsilon$. (b) Contour plots of the phase field $v$ over the undeformed configuration of the specimen right after fracture nucleation.}\label{Fig4}
\end{figure}
%
%

The $2^{\text{nd}}$ circle consists of a circular plate of radius $R = 5$~mm and thickness $B = 0.25$~mm that is subjected to equi-biaxial tension via a uniform radial displacement $u$ applied at its lateral boundary, while its top and bottom boundaries remain traction-free; see Fig.~\ref{Fig3}(a). This boundary condition generates a spatially uniform biaxial strain, $u/R$, and a corresponding uniform biaxial stress, $S$, within the plate. Precisely,
\begin{equation*}
\bfE=\dfrac{u}{R}(\bfe_1\otimes\bfe_1+\bfe_2\otimes\bfe_2)+e_t\bfe_3\otimes\bfe_3\quad {\rm and}\quad \bfS=S(\bfe_1\otimes\bfe_1+\bfe_2\otimes\bfe_2),
\end{equation*}
where $e_t$ denotes the transverse strain induced by the Poisson effect. This uniform state of stress and strain persists until the stress $S$ reaches the biaxial tensile strength $\sbs$ of the material. For our chosen silicate glass, this critical limit is $\sbs=2\sts\scs/(3\scs-\sts)=27$~MPa. Upon reaching this threshold, a through-thickness crack initiates instantly and sweeps across the specimen, fragmenting it. The stress $S$ and radial strain $u/R$ are therefore related by

\begin{equation}
S=\left\{\hspace{-0.15cm}\begin{array}{ll}
\dfrac{E u}{(1-\nu)R}& {\rm if}\;\; 0\leq u<u_{\texttt{bs}}\vspace{0.2cm}\\
0&{\rm if} \;\;u_{\texttt{bs}}\leq u\end{array}\right., \label{Response-Bi-Lin}
\end{equation}
where $u_{\texttt{bs}}$ denotes the value of the displacement at which $S=\sbs$. Due to the inherent spatial variations in strength, much like in the uniaxial tension test, the location of the severing crack is inherently arbitrary.

We exploit symmetry and model only a quarter of the plate. Figure~\ref{Fig4} presents the phase-field predictions. Specifically, part~(a) compares the stress-strain response ($S$ vs.~$u/R$) generated by the phase-field model for three distinct regularization lengths $\varepsilon$ against the reference result (\ref{Response-Bi-Lin}). Part~(b) displays the associated contour plots of the phase field $v$ over the undeformed configuration immediately following fracture nucleation. These findings show that as the regularization length $\varepsilon$ decreases, the phase-field model converges to the reference result.

\subsubsection{Torsion test}

%
\begin{figure}[t!]
\centering\includegraphics[width=0.775\linewidth]{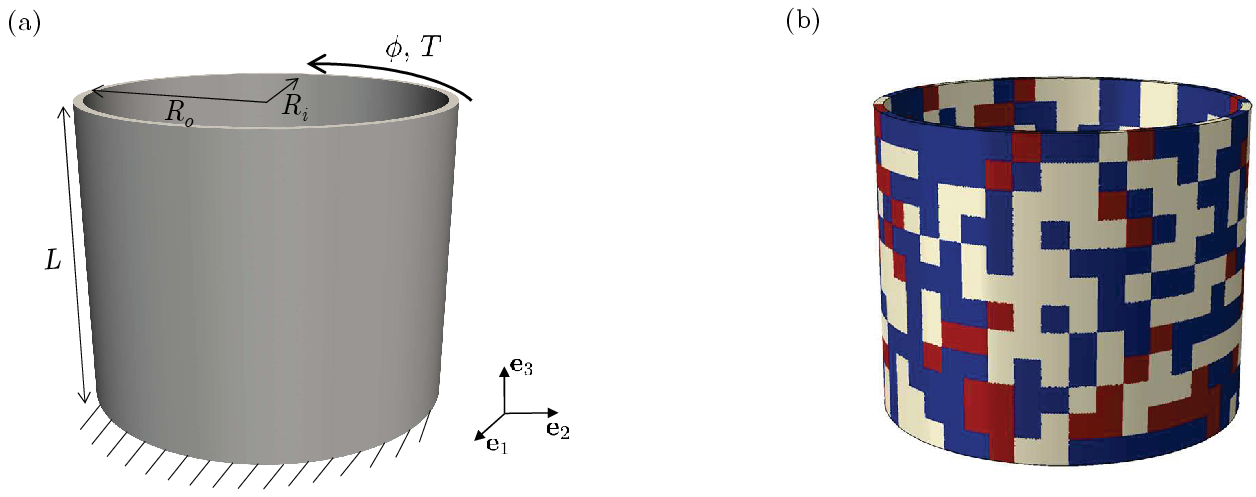}
\caption{\small (a) Schematic of the torsion test. The specimen dimensions are $L=5$ mm, $R_i=2.85$ mm, and $R_o=3$ mm. (b) A realization of the subregions of size $0.40$ mm over which the uniaxial tensile strength $\sts$ of the material is randomly perturbed.}\label{Fig5}
\end{figure}
%
%

Proceeding to the 3$^{\text{rd}}$ circle, we consider a thin-walled circular tube with an initial length of $L=5$ mm, an internal radius of $R_i=2.85$ mm, and an external radius of $R_o=3$ mm. The specimen undergoes torsion applied through a twist angle $\phi$ at one extremity, while the opposite end is held rigidly fixed, as illustrated in Fig.~\ref{Fig5}(a). This loading configuration induces a state of shear strain and shear stress within the tube of the form
\begin{equation}
\bfE=\dfrac{\phi X_1}{2L}(\bfe_2\otimes\bfe_3+\bfe_3\otimes\bfe_2)-\dfrac{\phi X_2}{2L}(\bfe_1\otimes\bfe_3+\bfe_3\otimes\bfe_1)=\gamma(\bfe_{\Theta}\otimes\bfe_Z+\bfe_Z\otimes\bfe_{\Theta})\label{E-gamma}
\end{equation}
and
\begin{align}
\bfS=\dfrac{2T X_1}{\pi(R_o^4-R_i^4)}(\bfe_2\otimes\bfe_3+\bfe_3\otimes\bfe_2)-\dfrac{2TX_2}{\pi(R_o^4-R_i^4)}(\bfe_1\otimes\bfe_3+\bfe_3\otimes\bfe_1)
=\tau(\bfe_{\Theta}\otimes\bfe_Z+\bfe_Z\otimes\bfe_{\Theta})\label{S-tau}
\end{align}
for $\bfX\in\Omega=\{\bfX:R_i^2<R^2<R_o^2,\,0<X_3<L\}$ with $R=\sqrt{X_1^2+X_2^2}$. Here, $\{\bfe_1,\bfe_2,\bfe_3\}$ is the laboratory frame depicted in the figure, whereas $\{\bfe_R,\bfe_{\Theta},\bfe_Z\}$ denotes the associated cylindrical frame of reference, $\gamma=\phi R/(2L)$ is the shear strain, $T$ denotes the resultant torque at the tube ends, and $\tau=2T R/(\pi(R_o^4-R_i^4))$ stands for the corresponding shear stress.

%
\begin{figure}[t!]
\centering\includegraphics[width=0.9\linewidth]{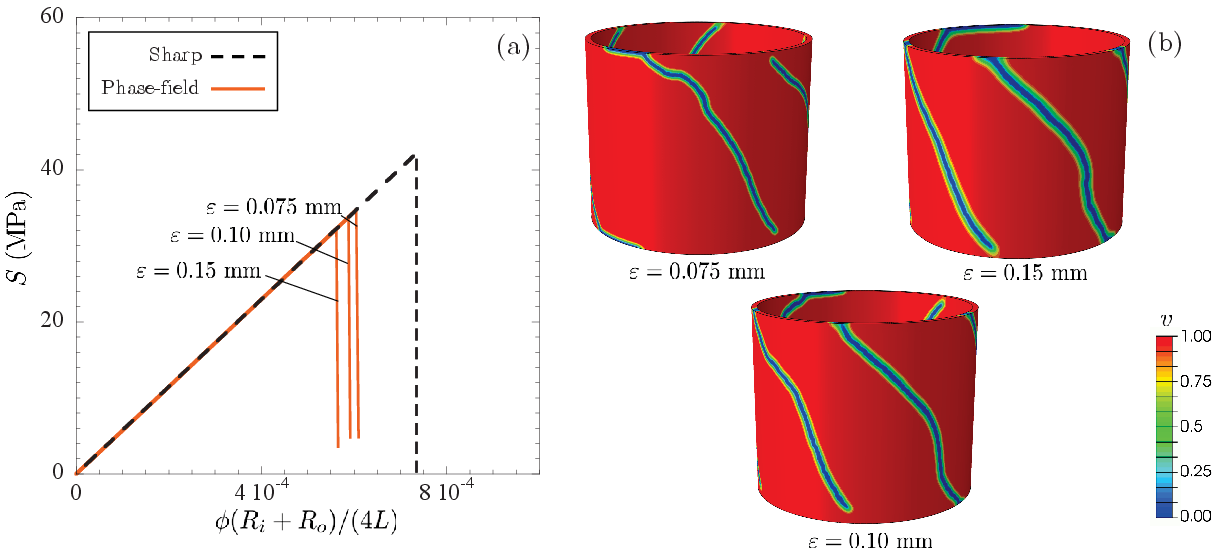}
\caption{\small (a) Comparisons between the predictions by the phase-field model for the stress-strain response of the tube and the reference result (\ref{Response-Shear}) for three different values of the regularization length $\varepsilon$. (b) Contour plots of the phase field $v$ over the undeformed configuration of the specimen right after fracture nucleation.}\label{Fig6}
\end{figure}
%
%

Given the thin-walled geometry of the tube ($t=R_o-R_i=0.15$ mm compared to $R_o=3$ mm), the corresponding stress (\ref{S-tau}) and strain (\ref{E-gamma}) are largely spatially uniform. They have average values of $S=T (R_i+R_o)/(\pi(R_o^4-R_i^4))=T(R_i-t)/(2\pi R_i^3 t)+O(t)$ and $\phi (R_i+R_o)/(4L)=\phi(2R_i+t)/(4L)$, respectively. This quasi-uniform state is maintained until the shear stress $S$ reaches the material's shear strength, $\sss$. For the particular silicate glass investigated here, $\sss=2\sts\scs/(\sqrt{3}(\scs+\sts))=44.4$~MPa. Upon reaching this critical limit, sudden fracture completely severs the specimen. Thus, the average stress and strain are related by
\begin{equation}
S=\left\{\hspace{-0.15cm}\begin{array}{ll}
\dfrac{E (R_i+R_o)\phi }{4(1+\nu)L}& {\rm if}\;\; 0\leq \phi<\phi_{\texttt{ss}}\vspace{0.2cm}\\
0&{\rm if} \;\;\phi_{\texttt{ss}}\leq \phi\end{array}\right.,  \label{Response-Shear}
\end{equation}
where $\phi_{\texttt{ss}}$ denotes the value of the angle of twist at which $S=\sss$. Although inherent material stochasticity dictates that these cracks initiate at random spatial locations, experimental observations demonstrate that they consistently exhibit a $45^\circ$ orientation relative to the symmetry axis $\bfe_3$.

Consistent with those in Figs.~\ref{Fig2} and \ref{Fig4}, the results in Fig.~\ref{Fig6} show once again that the phase-field model predicts accurately when and where fracture nucleates in the tube, provided that the value of the regularization length $\varepsilon$ is chosen to be sufficiently small. 

\subsection{Fracture nucleation dominated by Griffith energy competition}

\subsubsection{Pure-shear fracture test}

\begin{figure}[H]
\centering\includegraphics[width=0.9\linewidth]{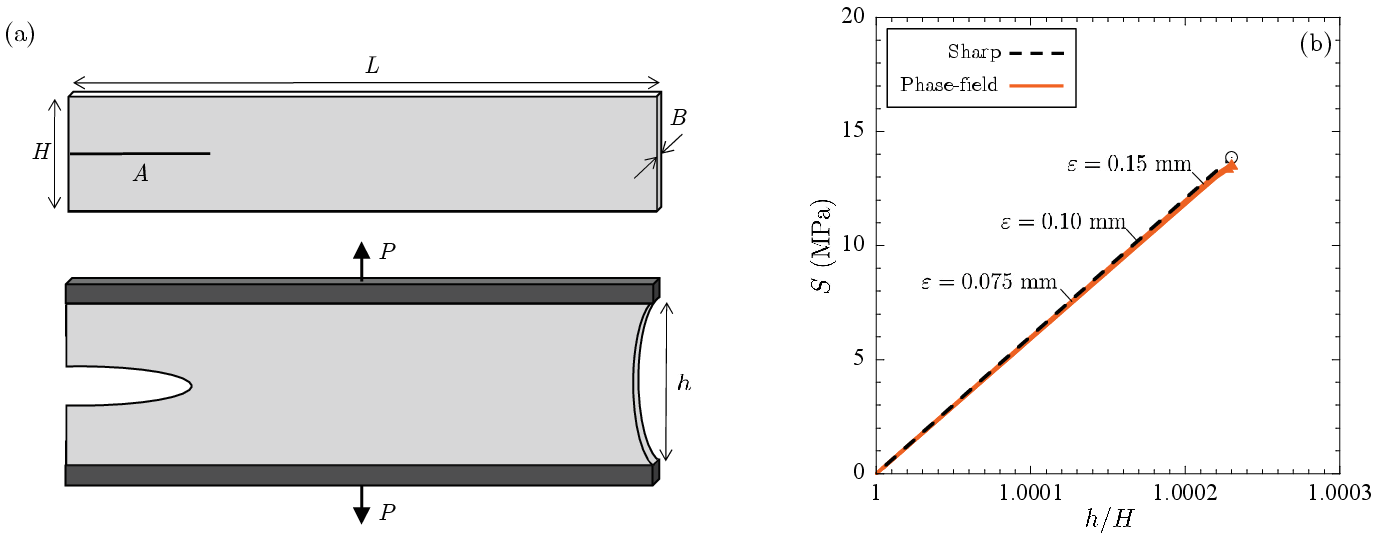}
\caption{\small (a) Schematic of the pure-shear fracture test in the undeformed and deformed configurations. The specimen dimensions are $L=50$~mm, $H=5$~mm, $B=0.5$~mm, and $A=10$~mm. (b) Phase-field predictions of the global response ($S=P/(L B)$ vs.~$h/H$) compared against the reference result for three regularization lengths $\varepsilon$. The curves terminate at the critical grip separation for fracture nucleation, given by (\ref{hcr-Lin}) for the reference result.}\label{Fig7}
\end{figure}

Figure~\ref{Fig7}(a) depicts the $4^{\text{th}}$ circle, a pure-shear fracture test on a thin strip of length $L=50$~mm, height $H=5$~mm, and thickness $B=0.5$~mm, featuring a pre-existing edge crack of initial length $A=10$~mm. The specimen is firmly clamped at its top and bottom boundaries and subjected to a prescribed separation $h$ between the grips. The resultant force at the grips is denoted by $P$. This classical strip geometry and loading allow one to determine a simple explicit estimate for the critical value $h_{cr}$ of the separation at which the crack will start to grow in a straight-ahead trajectory \cite{RT53}. The formula reads:
\begin{equation}
h_{cr}=\left(1+\sqrt{\dfrac{2(1-\nu^2)G_c}{HE}}\right)H.\label{hcr-Lin}
\end{equation}

Figure~\ref{Fig7}(b) compares the phase-field predictions of the global response, $S = P / (L B)$ vs.~$h / H$, for three regularization lengths $\varepsilon$, against the reference result. The simulations terminate at the critical grip separation corresponding to predicted fracture nucleation and demonstrate good agreement with the reference estimate (\ref{hcr-Lin}), irrespective of the value of $\varepsilon$.

%
%
%

\subsection{Fracture nucleation dominated by the interplay between strength and Griffith energy competition}

\subsubsection{Single edge notch test}

\begin{figure}[H]
\centering\includegraphics[width=0.9\linewidth]{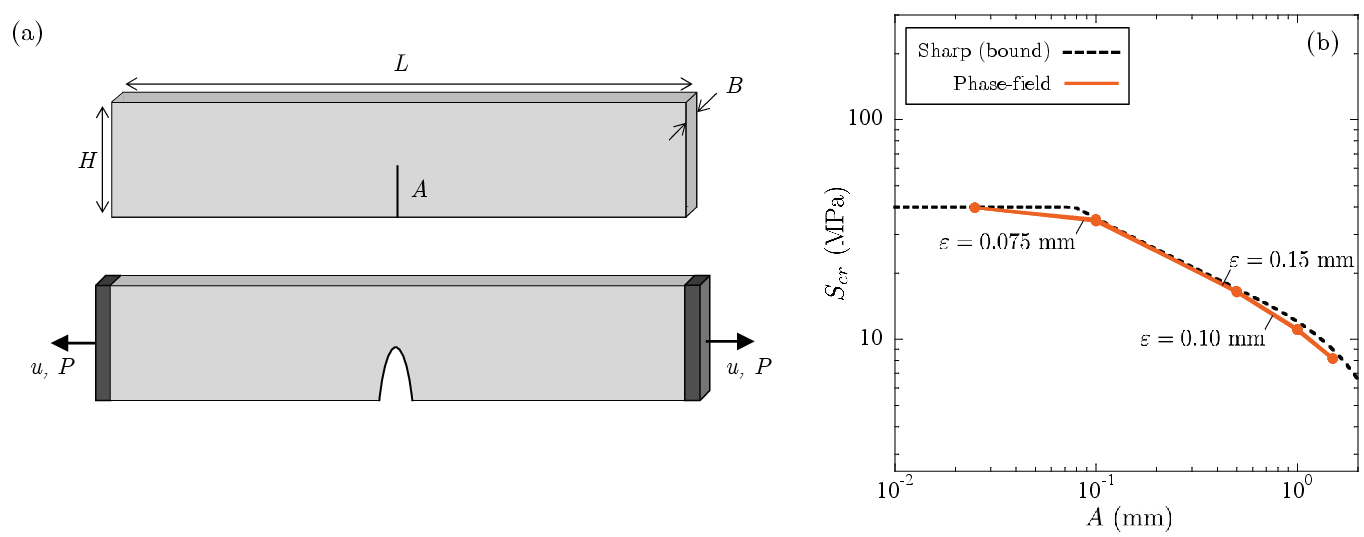}
\caption{\small (a) Schematic of the single edge notch test in the undeformed and deformed configurations. The specimen dimensions are $L=25$ mm, $H=5$ mm, $B=0.25$ mm, with pre-existing edge crack lengths $A=0.025, 0.1, 0.5, 1,$ and $1.5$ mm. (b) Comparisons between phase-field predictions for three values of the regularization length $\varepsilon$ and the bound (\ref{Scr-SENT-glass}) for the critical stress $S_{cr}$ at fracture nucleation. The results are shown as a function of the crack length $A$.}\label{Fig8}
\end{figure}

We proceed with the $5^{\text{th}}$ circle, the single edge notch test depicted in Fig.~\ref{Fig8}(a). The specimens are thin strips with a length $L = 25\text{ mm}$, a width $H = 5\text{ mm}$, and a thickness $B = 0.25\text{ mm}$. Each strip contains a pre-existing, through-thickness edge crack with an initial length $A$ selected from $0.025, 0.1, 0.5, 1,$ and $1.5\text{ mm}$. The left and right boundaries of the specimens are rigidly clamped and subjected to a prescribed global separation $u$, which induces a resultant reaction force $P$. The corresponding nominal global stress is defined as $S = P/(HB)$. Under increasing applied displacement $u$, the specimens exhibit an elastic response up to a critical threshold $u_{cr}$, at which point the pre-existing crack starts to grow in a straight-ahead trajectory. For the silicate glass under consideration, the critical global stress $S_{cr}$ at fracture nucleation is bounded from above  by the following relation \citep{KZDLP26}
\begin{equation}
S_{cr}\leq\min\left\{\sts,S_{G}\right\}\quad {\rm with}\quad S_{G}=\dfrac{\cos\left(\dfrac{\pi A}{2H}\right)\sqrt{\dfrac{G_c E}{\pi A}}}{\left(0.752+1.0431\dfrac{A}{H}+0.6076\left(1-\sin\left(\dfrac{\pi A}{2H}\right)\right)^3\right)\sqrt{\dfrac{2H}{\pi A}\tan\left(\dfrac{\pi A}{2H}\right)}}.\label{Scr-SENT-glass}
\end{equation}
In particular, the equality $S_{cr}=S_{G}$ is approached for sufficiently large cracks, while $S_{cr}=\sts$ is approached for sufficiently small cracks.

Figure \ref{Fig8} compares the phase-field predictions of the critical global stress $S_{cr}$, for three regularization lengths $\varepsilon$, against the reference bound (\ref{Scr-SENT-glass}). Consistent with all previous results, the predictions exhibit agreement with the reference result.

%
%
%

\subsubsection{Indentation test of a block with a flat-ended cylindrical punch}\label{Sec: Indentation}

%
\begin{figure}[t!]
\centering\includegraphics[width=0.9\linewidth]{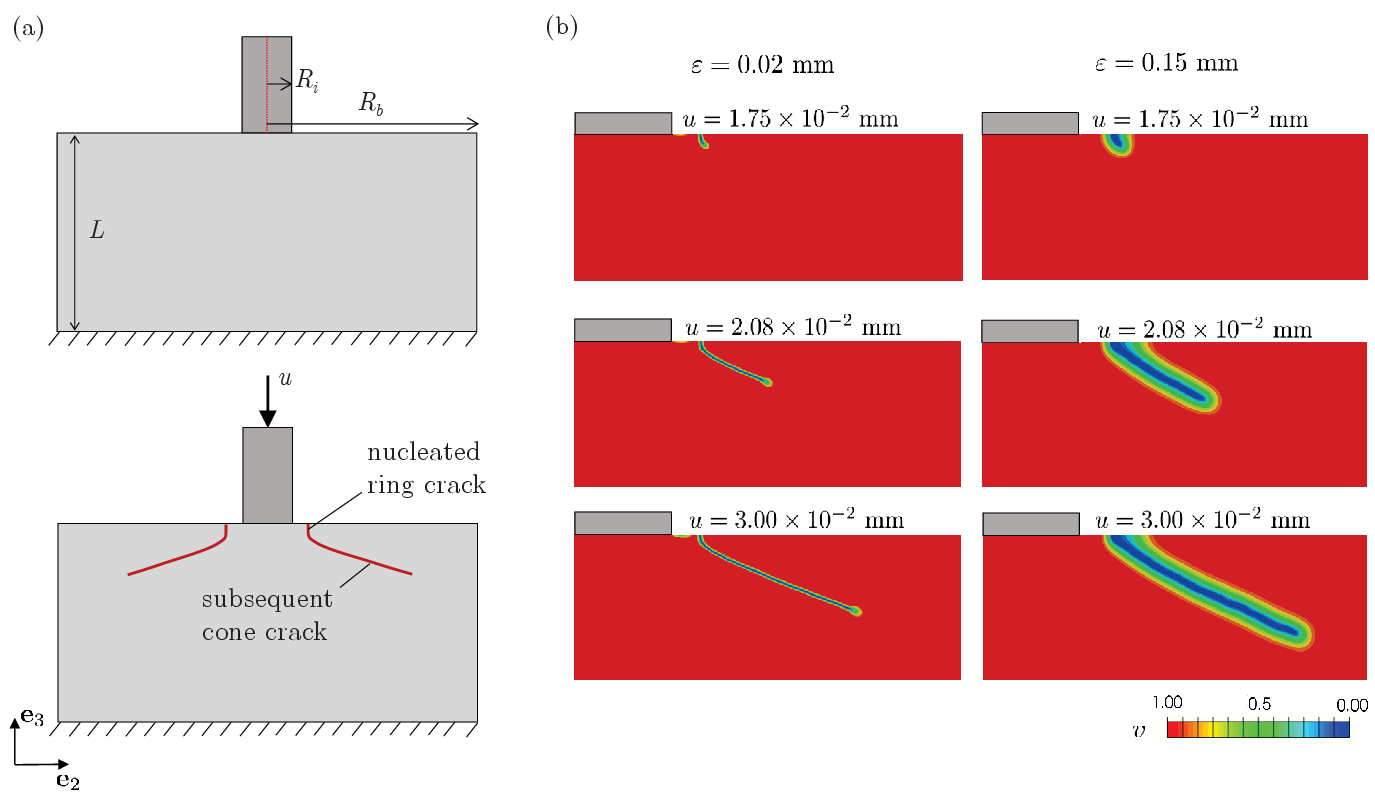}
\caption{\small (a) Schematic of the indentation test: a cylindrical block with height $L=25$ mm and radius $R_b=25$ mm, firmly bonded to a rigid substrate, that is frictionlessly indented along its axis of symmetry by a rigid, flat-ended cylindrical punch of radius $R_i=1$ mm under a prescribed displacement $u$. (b) Contour plots of the predicted phase field $v$ for two values of the regularization length $\varepsilon$ at three applied displacements $u$.}\label{Fig9}
\end{figure}
%
%

Figure \ref{Fig9}(a) depicts the setup for the $6^{\text{th}}$ circle. The specimen is a cylindrical block with a height $L=25$ mm and a radius $R_b=25$ mm, firmly bonded to a rigid substrate at its base. The block is frictionlessly indented along its axis of symmetry by a rigid, flat-ended cylindrical punch of radius $R_i=1$ mm under a prescribed displacement $u$. The silicate glass specimen initially exhibits a purely elastic response. Upon reaching a critical displacement threshold $u_{cr}$, a shallow ring crack nucleates on the free surface just outside the perimeter of the indenter ($R > R_i$). With continued loading, the crack propagates downward at a roughly constant angle relative to the surface, ultimately developing into a cone (frustum) crack.

Exploiting symmetry, we simulate this test as a 2D axisymmetric problem. Figure \ref{Fig9}(b) presents contour plots of the phase field $v$ over the undeformed configuration of the specimen, as predicted by the phase-field model for two regularization lengths $\varepsilon$. The plots correspond to three distinct applied indentation displacements $u$, with the first capturing the state immediately following fracture nucleation and the subsequent two at progressively larger displacements. As the regularization length $\varepsilon$ decreases, the numerical predictions converge to the reference result, capturing the nucleation of a ring crack and its subsequent evolution into a cone crack upon further loading.

\subsection{Fracture propagation}

\subsubsection{The double cantilever beam test}

%
\begin{figure}[b!]
\centering\includegraphics[width=0.99\linewidth]{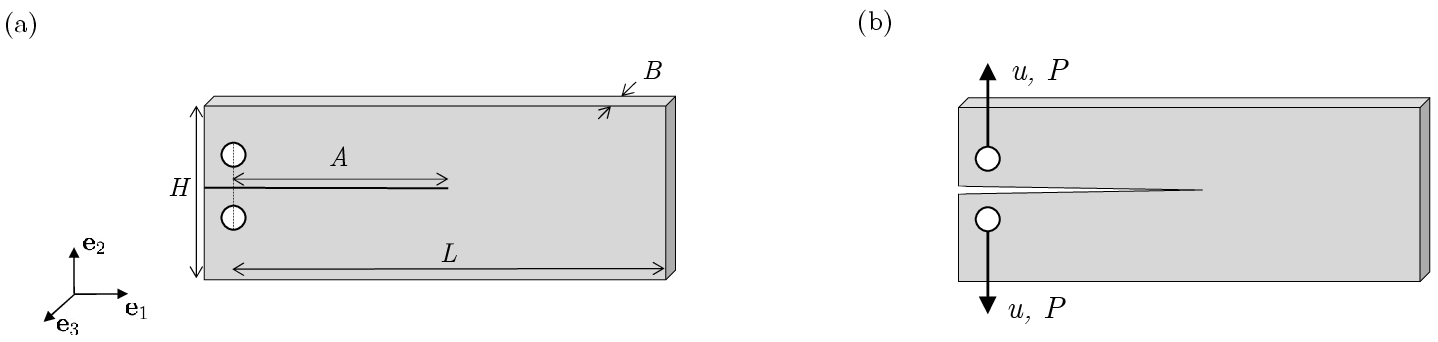}
\caption{\small Schematic of the double cantilever beam test. The specimen dimensions are $L=55$ mm, $H=20$ mm, $B=2.5$ mm, and $A=25$ mm. The radius of the pinholes is $B/8=0.3125$ mm and their centers are at a distance of $H/3=6.6667$ mm from each other and at a distance of $1.5$ mm from the left boundary of the specimen.}\label{Fig10}
\end{figure}
%
%

Finally, Fig.~\ref{Fig10} schematically illustrates the $8^{\text{th}}$ circle, a double cantilever beam test. The specimen comprises a prismatic bar measuring $L=55$ mm in length, $H=20$ mm in width, and $B=2.5$ mm in thickness, containing a pre-existing edge crack of initial length $A=25$ mm. Fracture is driven by a displacement $u$ applied along the $\bfe_2$ direction via two pinholes, generating a corresponding resultant force $P$. Each pinhole has a radius of $B/8=0.3125$ mm, is horizontally positioned $1.5$ mm inward from the left edge, and features a center-to-center vertical separation of $H/3=6.6667$ mm.
As the applied displacement $u$ increases, the specimen arms undergo flexure characteristic of cantilever beams. Upon reaching a critical displacement threshold $u_{cr}$, self-similar crack extension initiates in the $\bfe_1$ direction. Further increase in the applied displacement $u$ results in the stable propagation of the crack along the $\bfe_1$ direction.

%
\begin{figure}[t!]
\centering
\includegraphics[width=0.875\linewidth]{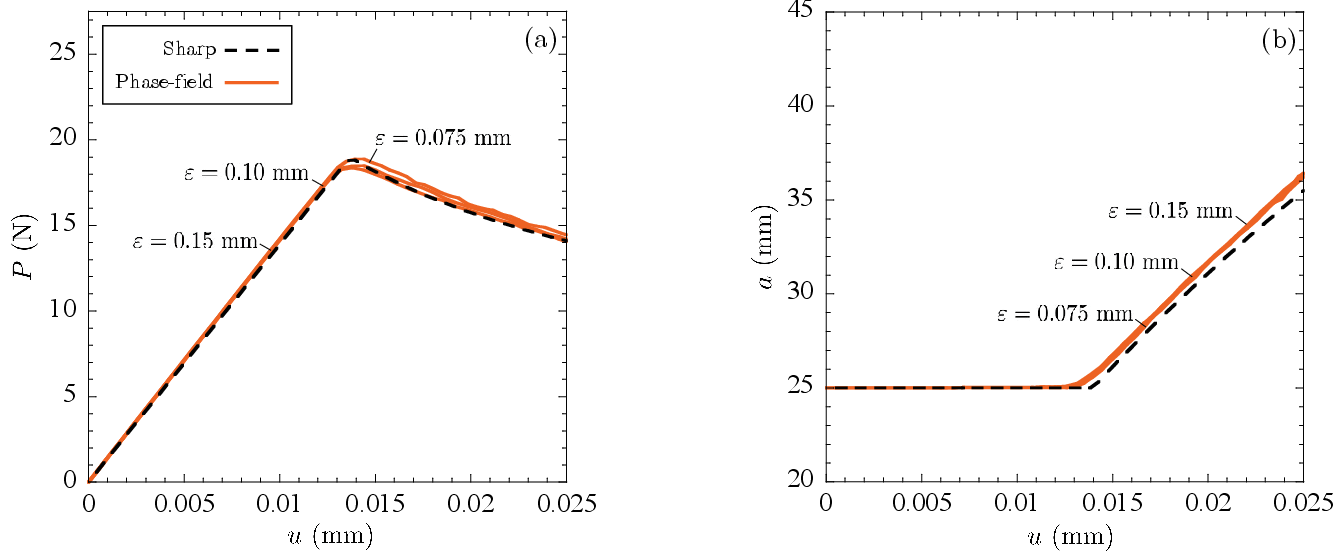}
\caption{{\small Phase-field predictions for the double cantilever beam test. (a) Comparisons between the predicted force-displacement response and the reference result (\ref{P-u-DCB}) for three values of the regularization length $\varepsilon$. (b) The associated evolution of the crack length $a$.}}\label{Fig11}
\end{figure}
%

As elaborated in \cite{KZDLP26}, this geometry and loading allow one to determine the explicit estimate for the global force-displacement response
\begin{equation}
P=\left\{\begin{array}{ll}
\dfrac{B H^{3+\alpha_1}\mu E}{32 H^{\alpha_1}\mu A^3+3\times 2^{3+\alpha_1}\alpha_2 H^2\mu A^{1+\alpha_1}+6 \alpha_3 H^{2+\alpha_1}E A}u &{\rm if}\;0\leq u< u_{cr}\vspace{0.2cm}\\
\dfrac{B H^{3+\alpha_1}\mu E}{32  H^{\alpha_1}\mu a^3+3\times 2^{3+\alpha_1}\alpha_2 H^2\mu a^{1+\alpha_1}+6  \alpha_3 H^{2+\alpha_1}E a}u  &{\rm if}\;u_{cr}\leq u\end{array}\right.\label{P-u-DCB}
\end{equation}
in terms of the explicit estimate 
\begin{equation}
a=\left\{\begin{array}{ll}
A &{\rm if}\;0\leq u< u_{cr}\vspace{0.2cm}\\
\textrm{Root closest to } A \textrm{ of the non-linear algebraic equation}\,g(a;u)=0  &{\rm if}\;u_{cr}\leq u\end{array}\right.,\label{a-DCB}
\end{equation}
for the evolving length $a$ of the crack, where
\begin{align*}
g(a;u)=\dfrac{2G_c a^2}{H^3\mu}-\dfrac{3H^{\alpha_1}E\left(16 H^{\alpha_1}\mu a^2+2^{2+\alpha_1}(1+\alpha_1)\alpha_2 H^2 \mu a^{\alpha_1}+\alpha_3 H^{2+\alpha_1}E\right)u^2}{\left(16 H^{\alpha_1}\mu a^2+3\times 2^{2+\alpha_1}\alpha_2 H^2\mu a^{\alpha_1}+3 \alpha_3 H^{2+\alpha_1}E\right)^2}.
\end{align*}
In these expressions, $\mu=E/(2(1+\nu))$ stands for the shear modulus and $\alpha_1$, $\alpha_2$, $\alpha_3$ are constants that exhibit small variations depending on how the boundary conditions are precisely applied. For the boundary conditions considered here, $\alpha_1=1.03$, $\alpha_2=0.658$, and $\alpha_3=0.70$.

Figure~\ref{Fig11}(a) compares the predicted force-displacement response ($P$ vs.~$u$) with the reference result (\ref{P-u-DCB}), while Fig.~\ref{Fig11}(b) plots the associated evolution of the crack length $a$ against (\ref{a-DCB}). Consistent with all previous findings, the phase-field model yields predictions that are in good agreement with the reference results.

Taken together, the above simulations demonstrate the capabilities and robustness of the proposed Abaqus user-subroutine implementation to study the nucleation and propagation of fracture in general boundary value problems involving structures made of isotropic linear elastic brittle materials.

\section*{Acknowledgements}

\noindent Support for this work by the National Science Foundation through the Grant DMS--2308169 is gratefully acknowledged.

\section*{Appendix. Visualizing the strength surface violation}

As emphasized in the Introduction, the violation of the strength surface ($\mathcal{F}(\bfS)\geq0$) is a necessary, but \emph{not} sufficient, condition for the nucleation and propagation of cracks. Identifying and visualizing the evolution of subdomains where $\mathcal{F}(\bfS)\geq0$ during a simulation thus provides direct insight into \emph{when} and \emph{where} cracks can potentially nucleate and propagate. As discussed in Section \ref{Sec: Output}, these subdomains are identified by the strength-violation indicator \texttt{SDV\_SSV\_IND}. In this appendix, by way of an example, we illustrate the evolution of \texttt{SDV\_SSV\_IND} for the indentation problem presented in Section \ref{Sec: Indentation}.

\begin{figure}[b!]
\centering\includegraphics[width=0.75\linewidth]{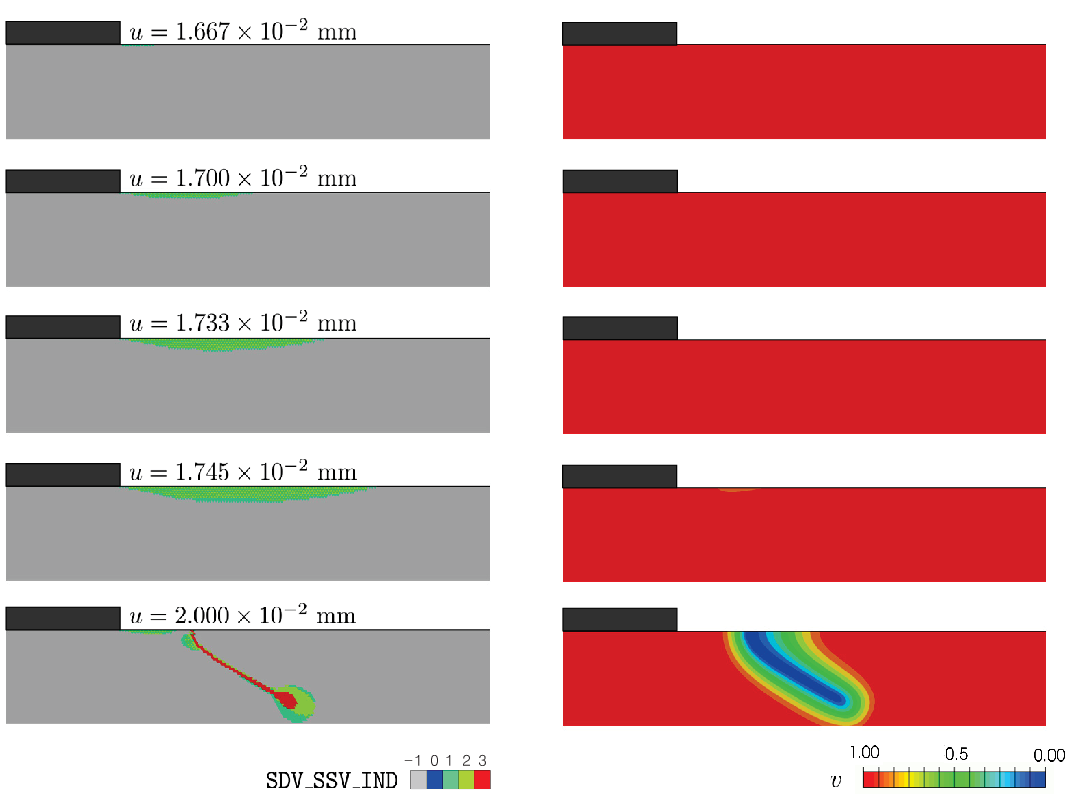}
\caption{\small Contour plots of the strength-violation indicator \texttt{SDV\_SSV\_IND} (left) and the phase field $v$ (right) near the flat-ended punch for the indentation test of Section \ref{Sec: Indentation} with $\varepsilon = 0.15\text{ mm}$, shown at five increasing applied displacements $u$.}\label{Fig12}
\end{figure}

Figure \ref{Fig12} shows side-by-side contour plots of \texttt{SDV\_SSV\_IND} and the corresponding phase field $v$ (\texttt{NT11}) over a magnified region near the flat-ended cylindrical punch for five increasing values of the applied displacement $u$. The results pertain to the regularization length $\varepsilon=0.15$ mm. At small indentation depths (top row), the strength surface is already violated ($\texttt{SDV\_SSV\_IND} = 2$) within a small ring-shaped subdomain on the free surface right outside the perimeter of the punch, yet the material remains fully intact ($v = 1$). As the applied displacement $u$ increases (second through fourth rows), this zone of strength violation expands radially outward and downward into the bulk while maintaining \texttt{SDV\_SSV\_IND} $= 2$. Crucially, even though the material strength is exceeded over a growing finite volume in rows one through four, no crack nucleates yet ($v \approx 1$). Once the applied displacement reaches the critical threshold $u_{cr}\approx 1.75\times 10^{-2}$ mm, it becomes energetically favorable to nucleate a crack ($v = 0$) within the strength-violating subdomain. In the post-nucleation regime ($u>u_{cr}$), the stress state in the wake of the formed crack relaxes back inside the strength surface ($\texttt{SDV\_SSV\_IND} = -1$), while the active zone of strength violation localizes around the propagating crack tip, where high stress triaxiality is evidenced by a core of three strictly positive principal stresses ($\texttt{SDV\_SSV\_IND} = 3$) surrounded by a halo of $\texttt{SDV\_SSV\_IND} = 2$ and $1$.

%
%
%

\end{document}